\documentclass[preprint2]{aastex}
\usepackage{natbib}
\usepackage{graphicx}
\newcommand{\xray}{X--ray}
\newcommand{\xmm}{{\it XMM-Newton}}
\newcommand{\rgs}{{\it{RGS}}}
\newcommand{\chandra}{{\it{Chandra}}}
\newcommand{\letgs}{{\it{LETGS}}}
\newcommand{\mrk}{Mrk~421}

\shorttitle{\xray Absorption by $z>0$ \ion{O}{vii} toward \mrk}
\shortauthors{Rasmussen et al.}

\begin{document}

\title{On the Putative Detection of $z>0$ \xray\ Absorption Features \\
  in the Spectrum of \mrk}

\author{Andrew P. Rasmussen\altaffilmark{1}, Steven
  M. Kahn\altaffilmark{1,2}}
\affil{Stanford Linear Accelerator Center, Menlo Park, CA 94025}
\email{arasmus@slac.stanford.edu}
\author{Frits Paerels\altaffilmark{3}}
\affil{Columbia Astrophysics Laboratory, New York, NY 10027}
\author{Jan~Willem~den~Herder, Jelle~Kaastra and Cor~de~Vries}
\affil{S.R.O.N., Sorbonnelaan 2, 3584 CA Utrecht, The Netherlands}

\altaffiltext{1}{Kavli Institute for Particle Astrophysics and
  Cosmology, Stanford University}
\altaffiltext{2}{Stanford University Department of Physics}
\altaffiltext{3}{Astronomy Department, Columbia University}

\begin{abstract}
  In a series of papers, Nicastro et al. have 
%  claimed 
  reported
  the detection of
  $z>0$ \ion{O}{7} absorption features in the spectrum of \mrk\
  obtained with the \chandra\ Low Energy Transmission Grating
  Spectrometer (\letgs). We evaluate 
%  those claims 
  this result
  in the context of a
  high quality spectrum of the same source obtained with the
  Reflection Grating Spectrometer (\rgs) on \xmm. 
  The data 
  comprise over 955~ksec of usable exposure time and more than
  $2.6\times 10^4$ counts per 50~m\AA\ at 21.6\AA. 
  We concentrate 
  on the spectrally clean region ($21.3 < \lambda < 22.5$\AA)
  where sharp features due to the astrophysically abundant \ion{O}{7}
  may reveal an intervening, warm--hot intergalactic medium (WHIM).
%  In spite of 
%  the fact that the sensitivity of the \rgs\ data is higher than that
%  of the original \letgs\ data presented by Nicastro et al., 
%  we do not confirm
  We do not confirm
  detection of any of the intervening systems claimed to date. Rather,
  we detect only three unsurprising,  {\it astrophysically expected}
  features down to the Log$\left({N_i}\right)\sim 14.6$ ($3\sigma$)
  sensitivity level. 
  Each of the two purported WHIM 
  features is rejected 
  with a statistical confidence that exceeds that reported for its
  initial detection.
  While we can not rule out the existence of fainter, WHIM
  related features in these spectra, we suggest that previous
  discovery claims were premature. A more recent paper by Williams et
  al. claims to have demonstrated that the \rgs\ data we analyze here
  do not have the resolution or statistical quality required to
  confirm or deny the \letgs\ detections. 
  We show that our careful analysis 
%of the \rgs\ data yields different results. 
  resolves the issues encountered by Williams et al. and recovers the
  full resolution and statistical quality of the RGS data.
  We highlight the differences between our analysis and those
  published by Williams et al. as this may explain 
  our disparate conclusions.
% JWDH:
%  differences in the
%  conclusions drawn.
% original
%   We show that the Williams et
%   al. reduction of the \rgs\ data was highly flawed, leading to an
%   artificial and spurious degradation of the instrument response. We
%   carefully highlight the differences between our analysis presented
%   here and those published by Williams et al.
\end{abstract}


\section{Introduction}
With the advent of high resolution X-ray spectroscopy provided by the
diffraction grating spectrometers on the {\it Chandra} and {\it XMM-Newton}
observatories \citep{Canizares00,Brinkman00,denHerder01}, it has become feasible to undertake an exploratory search
for  the large amount of baryonic matter that may be contained in a highly
ionized phase of the  local Intergalactic Medium. The agreement of the baryon
density at high redshift as measured in the  Ly$\alpha$ forest, compared to the
density predicted from Big Bang  Nucleosynthesis and the light element
abundances \citep{Cowie95,Burles97,Burles98} strongly suggests that such a medium should exist.
Independently,  large scale coupled dark matter/hydrodynamics simulations have
shown that an early, largely neutral IGM will progressively become more highly
ionized, until at the present day it is essentially undetectable at optical/UV
wavelengths \citep{Cen99,Croft01}. The calculations indicate that a major fraction of the
baryons at small redshift could reside in this diffuse, warm,
unvirialized phase of the IGM, and this is consistent with the fact
that the local baryon density 
inferred from a census of  stars and gas in virialized structures
falls short of the predicted value by up to 50\%.

Currently, the most promising technique for detecting and characterizing the
medium is high resolution soft X-ray absorption spectroscopy of the low-$Z$
elements towards bright extragalactic continuum sources, which has now been
attempted  with a variety of instruments towards a number of suitably
bright, spectrally featureless objects
\citep{Fang00,Fang01,FBC02,Fang02,Nicastro02,Rasmussen03a,Rasmussen03b,Mathur03,Cagnoni04,Ravasio05,Nicastro05,Barcons05}. 
The search has naturally focused on the K shell resonance lines of H- and
He-like oxygen (\ion{O}{7} $1s^2\ ^1S_0-1s\,2p\ ^1P_1$ ($w$), 21.602~\AA,
and \ion{O}{8} 
$1s-2p$, 18.969~\AA), since oxygen has high abundance,
and the O K band is relatively clean.  Thus far, these observations have not
produced an unambiguous detection of highly ionized metals except at zero
redshift, where resonance absorption in C, O, and Ne is associated
with gas in and around the Milky Way Galaxy, and possibly in a tenuous
intragroup medium in the Local group. This state of affairs is not surprising,
given the predicted distribution of column densities \citep{FBC02,Chen03},
the fact that only a relatively small redshift path has been surveyed, and 
that the wavelength resolution of the grating spectrometers on {\it
  Chandra} and{\it XMM-Newton} is roughly an order of magnitude
too poor to provide adequate sensitivity for the expected equivalent
widths. 

The most ambitious and most promising search has been conducted using a very
deep spectrum of the  blazar Mrk 421, obtained with the Low Energy Transmission
Grating Spectrometer (\letgs) on {\it Chandra} (using  both the ACIS-S and HRC-S cameras), 
collecting data taken during times when the source was undergoing an outburst
\citep{Nicastro05N,Nicastro05}. The most surprising feature in this
spectrum is the presence of what appear to be faint  
\ion{O}{7}~$w$ (21.602~\AA) 
resonance absorption lines, at redshifts $z \approx 0.011$ and 
$z \approx 0.027$ (the redshift of Mrk 421 \citep{Ulrich75} is $z
\approx 0.0308$), and additional absorption, though much weaker, in
other transitions at approximately the same redshifts.  
These reported detections have attracted substantial attention, since,
if correct, they would represent the first discovery of the hottest
phase of the WHIM, long predicted by cosmological N-body simulations.

We have analyzed an even deeper spectrum based on data obtained with the
Reflection Grating Spectrometer (\rgs) on {\it XMM-Newton}, comprising nearly a
million seconds of exposure time, in a spectrometer with approximately
3 times the effective area of {\it LETG/HRC-S} ($5-8$ times the effective
area of {\it LETG/ACIS-S}) in the relevant spectral band, at comparable
wavelength resolution. We do not confirm the detectiion of the $z>0$
\ion{O}{7} absorption features reported by \citet{Nicastro05N,Nicastro05}.

In a recent paper, \citet{Williams06} report on the analysis of a
portion of these same \rgs\ observations and claim to have shown that
the data are not of sufficient quality to confirm or deny the
\chandra\ detections. 
On the contrary: We show that a careful reduction of the \rgs\ data 
does not suffer from the serious limitations cited by them, and that
the \rgs\ instrument may have been 
% unduly 
incorrectly 
blamed for the shortcomings
of their attempted analysis.
% original
%  We show here that \citet{Williams06} invoke a
%  faulty analysis which artificially degraded the \rgs\ response.

In the following, we describe the RGS dataset on Mrk 421 and we give a detailed
description of our data analysis procedures. We then describe the search for
faint, redshifted discrete absorption lines in the RGS spectrum, and show that
no significant redshifted absorption is seen in the O~K band. We quantify our
result that absorption lines at the contrast seen in \letgs\ can be ruled
out. 
We compare our analysis with that of \citet{Williams06} and illustrate
what we believe the differences are.
% original
%  where they went wrong.
We conclude with a summary, and briefly put our finding in the context
of the search for the  highly ionized IGM.

\section{Observations}
Already in its sixth year in orbit, \xmm\ has pointed toward \mrk\
many times for various purposes. Our strategy to maximize our
sensitivity for detecting faint spectral features logically requires
combining as much data as possible, drawing from the multiple
observations that are available. Table~\ref{tab:mrk_obs_table}
summarizes the 33 \xmm\ pointings (out of 36 currently available) that
we used in our analysis. We combined data from these observation data
files (ODF) into 14 different data sets. After rejecting periods of
high background, the remaining ``good'' integration time
makes up roughly 93\% of the total (exceeding 1~Ms), at 955~ks. 
Three ODFs that were neglected would have contributed only 10~ks in
additional exposure.  

\begin{deluxetable}{cccc}
\tablecolumns{4}
\tablewidth{0pc}
\tablecaption{Mrk~421 observations for absorption analysis.\label{tab:mrk_obs_table}}
\tablehead{
\colhead{XMM--Newton } &
\colhead{Obs. Start} & 
\colhead{Obs. On} & 
\colhead{Target} \\
\colhead{Obs. ID} & 
\colhead{Date (UT)} & 
\colhead{Time [s]} & 
\colhead{Dataset} 
}
\startdata
0099280101 & 2000-05-25 & 66497       & 0084 \\
0099280201 & 2000-11-01 & 40115       & 0165 \\
0099280301 & 2000-11-13 & 49811       & 0171 \\
0099280401 & 2000-11-14 & 43010       & 0171 \\
0099280501 & 2000-11-13 & 21206       & 0171 \\
0099280601 & 2000-11-15 & 20213       & 0171 \\
0136540101 & 2001-05-08 & 39007       & 0259 \\
0136540201 & 2001-05-08 & \phn9816    & 0259 \\
0153950601 & 2002-05-04 & 39727       & 0440 \\
0153950701 & 2002-05-05 & 19982       & 0440 \\
0153950801 & 2002-05-05 & 21671       & 0440 \\
0136540301 & 2002-11-04 & 23913       & 0532 \\
0136540401\tablenotemark{a} & 2002-11-04 & 23917       & 0532 \\
0136540501\tablenotemark{a} & 2002-11-04 & 22914       & 0532 \\
0136540601\tablenotemark{a} & 2002-11-04 & 22917       & 0532 \\
0155555501\tablenotemark{a} & 2002-11-05 & 37765       & 0532 \\
0136540701\tablenotemark{b} & 2002-11-14 & 71520       & 0537 \\
0136540801\tablenotemark{b} & 2002-11-14 & 11415       & 0537 \\
0136540901\tablenotemark{b} & 2002-11-15 & 11420       & 0537 \\
0136541001 & 2002-12-01 & 71118       & 0546 \\
0136541101 & 2002-12-02 & 11413       & 0546 \\
0136541201 & 2002-12-02 & 11415       & 0546 \\
0158970101 & 2003-06-01 & 47538       & 0637 \\
0158970201 & 2003-06-02 & \phn8963    & 0637 \\
0158970701 & 2003-06-07 & \phn8055    & 0640 \\
0158970801 & 2003-06-07 & 12805       & 0640 \\
0158970901 & 2003-06-08 & 10752       & 0640 \\
0158971001 & 2003-06-08 & 12800       & 0640 \\
0150498701 & 2003-11-14 & 48917       & 0720 \\
0162960101 & 2003-12-10 & 39889       & 0733 \\
0158971201 & 2004-05-06 & 66141       & 0807 \\
0153951201 & 2005-11-07 & 10017       & 108x \\
0158971301 & 2005-11-09 & 60015       & 108x \\
\enddata
\tablenotetext{a}{RFC (RGS2) Cooldown}
\tablenotetext{b}{RFC (RGS1) Cooldown}
\end{deluxetable}

The data sets that were analyzed are summarized in
Table~\ref{tab:mrk_dataset_table}. Essential data products (spectra
and response matrices) for each data set were prepared using filtering
parameters specific to the content of the constituent ODFs. A
description of the filtering specifications is included below.

\begin{deluxetable}{ccccccr}
\tablecolumns{7}
\tablewidth{0pc}
\tablecaption{Spectral data sets produced.\label{tab:mrk_dataset_table}}
\tablehead{
  \colhead{Dataset} &
  \colhead{GTI} & 
  \colhead{${N \over \Delta\lambda}$\tablenotemark{a}} &
  \colhead{${\Delta R_{90} \over R}$\tablenotemark{b}} &
  \colhead{$\Delta \phi_{90}$\tablenotemark{b}} &
  \colhead{$S$\tablenotemark{b}} &
  \colhead{$\chi^2_\nu$\tablenotemark{c}}}
\startdata
0084                    & \phn63.52  &  \phn1670    & 0.29 &  0.004 &  0.001  & 0.912  \\
0165                    & \phn36.34  &  \phn\phn426 & 0.33 &  0.010 &  0.003  & 0.900  \\
0171                    &    133.03  &  \phn4383    & 0.25 &  0.241 &  0.061  & 1.197  \\
0259                    & \phn40.84  &  \phn1007    & 0.23 &  0.010 &  0.002  & 1.042  \\
0440                    & \phn80.86  &  \phn\phn862 & 0.62 &  0.107 &  0.067  & 0.979  \\
0532                    & \phn92.31  &  \phn3044    & 0.52 &  0.050 &  0.026  & 1.331  \\
0537                     & \phn91.04  &  \phn3028    & 0.48 &  0.169 &  0.081  & 1.028  \\
0546                    & \phn93.66  &  \phn1722    & 0.29 &  0.168 &  0.049  & 1.000  \\
0637                    & \phn62.95  &  \phn1276    & 0.26 &  0.104 &  0.027  & 1.051  \\
0640                    & \phn49.02  &  \phn\phn584 & 0.18 &  0.517 &  0.096  & 1.194  \\
0720                    & \phn48.80  &  \phn2094    & 0.28 &  0.024 &  0.007  & 1.054  \\
0733                    & \phn27.78  &  \phn\phn572 & 0.17 &  0.012 &  0.002  & 0.923  \\
0807                    & \phn65.98  &  \phn2632    & 0.28 &  0.006 &  0.002  & 1.222  \\
108x                    & \phn69.82  &  \phn2888    & 0.23 &  0.102 &  0.023  & 0.770  \\
\hline
\multicolumn{1}{l}{Total}      & 955.90      & 26188 &\nodata&\nodata&\nodata&1.040 \\
\enddata
\tablenotetext{a}{counts per 50~m\AA\ spectral range at 21.6\AA}
\tablenotetext{b}{90\% widths of the relative RGS count rate distribution, off-axis angle distribution, and their product $S\equiv\Delta\phi_{90}\times{\Delta R_{90}\over R}$} 
\tablenotetext{c}{fits to the $\Delta\lambda$=1.2\AA\ band, with 7.4~m\AA\ bins} 
\end{deluxetable}

The usual data analysis paradigm that applies to our approach is one
where a data set's integration time occurs over a period where the
source and all instrument characteristics are {\it assumed} to be
static. 
The reality is substantially different, particularly for long
observations. Even in a single observation, many parameters of an
observation change, including the source's
spectrum, the satellite's attitude, and the number and location of
problematic detector areas that are identified and subsequently
rejected from analysis. Variation of any of these quantities can lead
to subtle complications and may frustrate simple interpretation of
weak features in the data. The proper treatment of these effects is
detailed below. 

\section{Data Analysis}

The \rgs\ branch of the \xmm\ Science Analysis System
(SAS\footnote{http://xmm.vilspa.esa.es/external/xmm\_sw\_cal/sas.shtml}) 
provides software filtering for \xray\ events detected in the readout
CCDs. It is designed to provide nominal data products essential for
spectral modeling (spectra, background samples and response matrices)
under the assumption that the \xray\ event data can be converted into
a form where the instrumental signature of each event's origin has
been removed. 
While the response matrix generator can provide proper compensation
for physical QE variation across the CCD array, it specifically
does not compensate for QE variations due to local charge transport
anomalies across the 18 detector readouts per \rgs. 
As the CCDs incur more radiation damage, this becomes a limitation of
the existing analysis approach. Thus, while overall charge transport
characteristics are corrected for, the event ``pulse--invariant''
conversion model does not include detail on the pixel scale.
Currently, only four charge transfer inefficiency (CTI) parameters per
readout node are used to perform this correction.

We find that the majority of faint, systematics--induced 
features arise from three mechanisms: 
\begin{enumerate}
\item{Localized gain anomalies (attributed to radiation damage induced
  CTI variations). These are identified and corrected for in hardware
  coordinates. In general these are limited to a small fraction
  ($<$1\%) of CCD area, particularly after camera cooldown was
  performed circa rev.~0530.}  

\item{Transient, high duty--cycle pixel reads. These can evade
  identification in long exposure times, where the hot pixel finder
  will succeed in finding persistent pixels even at a low duty cycle.}

\item{Crosstalk pixels -- pickup of synchronously sampled analog
  signal of high dark current pixels. The CCDs are each read out of
  two output amplifiers and ``mirror images'' of cosmic rays are seen
  in the electronic image of the other readout, at a $\sim$1-2\%
  crosstalk coupling. }

\item{Changes in source spectrum in the presence of finite
  spacecraft drift. Systematics are introduced in regions of
  non-continuous detector coverage. The effect is present only if both
  spectral variation {\it and} spacecraft drifts occur within an
  observation. It is analogous to the effect of adding 
  spectra from multiple offset pointings where the source varied
  between pointings ({\it cf.} \S~\ref{comp_to_williams}).}

\end{enumerate}
The first three mechanisms listed above have been observed with varying
degrees of significance, and each provides a mediating process for 
redistributing nominal \xray\ induced CCD events out of the event
extraction pulseheight window (thereby inducing faint systematics that
can resemble absorption features). 
The fourth mechanism can affect the interpretation of the data in the
context of the common approach to analysis\footnote{Consider
  co--adding raw spectral data from two 
  phases ({\it e.g.}, dim and bright) of a source spectrum, where gaps
  in spectral coverage have also moved between the phases. When a
  single exposure map is used in the analysis, bimodal features are
  induced in the residuals to any spectral fit, whose locations
  correspond to the gap's position in each of the two phases.}.
Three columns in Table~\ref{tab:mrk_dataset_table} address some of
these effects (${\Delta R_{90}  \over R}$, $\Delta\phi_{90}$ \& $S$). 
These correspond to the 90\% distribution widths in the relative
countrate and the pointing variation (in arcminutes) and their product
($S\equiv \Delta\phi_{90}\times{\Delta R_{90} \over R}$), respectively.
Contribution to systematics by this mechanism may scale with a data
set's susceptibility $S$ and the number of affected regions
should scale with the number of detector regions excluded from the
analysis.

Our production of data sets that are minimally impacted by
systematics focuses on identification and rejection of problematic
detector regions and is guided by the following principles: 
\begin{enumerate}
\item{The effective exposure time of each data set should be
  sufficient to generate adequate statistics for a quiescent
  thresholded hot pixel map and median offset map for each
  CCD. Conversely, the instrumental detail for each data set should be
  specific to its epoch and not averaged over other observations where
  detector characteristics may have changed. We have 
  chosen to combine ODF data that span approximately full \xmm\
  revolutions ($\sim$48 hours).} 
\item{An iterative approach is adopted for rejection of
  problematic detector areas. Automatic detection and
  rejection of some regions is performed by simultaneously
  inspecting both a median offset CCD frame and a thresholded hot
  pixel map for each CCD analyzed. A second screening is performed by
  manual inspection of the surviving \xray\ events, arranged by
  hardware and pulse--invariant event parameters. Non--statistical
  features are readily identified when superior counting statistics
  are available. An example for the iterative approach to data
  screening is provided in the Appendix.}
\end{enumerate}
While we used custom software to prepare the datasets for spectral
analysis, we emphasize that if used thoughtfully, {\it any} software
can be used to achieve these results. In the Appendix we provide a
demonstration that the \xmm\ SAS can be used (interactively and
iteratively) to produce an equivalent spectrum. We cannot
guarantee this if one were to start with the data products 
automatically generated by the SAS pipeline processing system
(PPS). This limitation is consistent with the purpose of the PPS
products: They are intended to provide the user with a ``quick--look''
assessment of the ODF's contents, and no attempt is made in their
preparation to reduce systematics.
% original 
%  beyond the most basic level.

\begin{figure}
\includegraphics[angle=-90,width=\columnwidth]{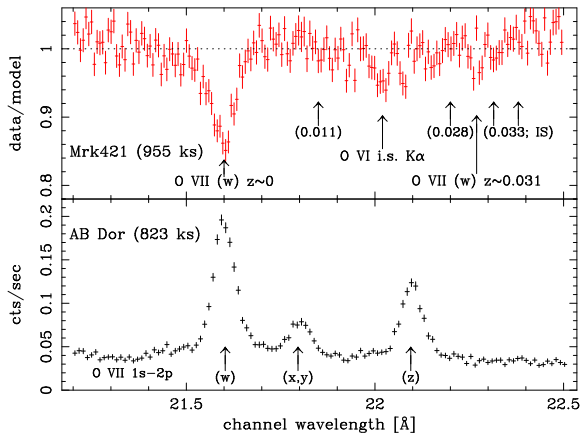}
\caption{\label{fig:mrk421_resids}
  The deep (955~ksec) absorption spectrum toward 
  \mrk\ (top panel). Locations for intervening absorption features
  previously reported by   \citet{Nicastro05} are indicated with
  parenthesized redshifts. The \mrk\ spectrum is given alongside a
  line--rich coronal spectrum (bottom panel) of comparable integration
  time and observation count for comparison. Both plots use the
  instrument's {\it blind wavelength scale} for their constituent
  datasets. The \ion{O}{7}~$w$ emission line profile in the spectrum
  of AB~Dor provides a model- and calibration-independent estimate for
  the effective wavelength response in spectra composed of many
  different datasets. }
\end{figure}

Each data set generated was then analyzed in parallel within XSPEC
\citep{Arnaud96} as a separate spectrum paired with its specific
response matrix and backlight emission model. 
A total of 10 isolated channels (out of a total 2268) stood out by
contributing $\Delta\chi^2>10$ and were excluded from the
fit. Indeed, 8 of these 10 channels reside in the 6 datasets
with the susceptibility parameter $S>0.025$\arcmin\ ({\it cf.}
Tab.~\ref{tab:mrk_dataset_table}), consistent with some of our initial 
considerations on systematic errors.   

The spectra of all the data sets included in the joint fit are
combined only for plotting purposes via XSPEC's {\tt setplot group}
command. The resulting absorption spectrum is given in 
Figure~\ref{fig:mrk421_resids}, where each spectrum was divided
through by its specific folded powerlaw continuum model prior to
averaging across data sets. In the following section we discuss the
quantitative analysis of this composite spectrum representing 955~ks
of integration time toward \mrk.

\section{Analysis of the 21.2--22.5\AA\ spectrum}
\subsection{Continuum modeling; the discrete absorption at redshift zero}

The absorption spectrum in the range where the \ion{O}{7}~$w$ line should
appear for the redshift range $z = 0-0.030$ was generated as described
above.
No additional splines to the
instrument response were applied, so any uncorrected, differential calibration
errors over this spectral range will appear in these residuals.  We neglected
any absorption by neutral gas in our Galaxy in the fitting, since our analysis
spans a narrow wavelength band. The foreground column density is small
($N_{\mathrm H} = 1.61 \times 10^{20}$ cm$^{-2}$, Lockman \& Savage 1995), and
the transmission $T$ of the ISM varies by 2\% across the chosen range,
but can be approximated locally by a power law $T\sim\lambda^{-0.38}$
within 0.02~\% 
accuracy. The most prominent absorption features ($\lambda=21.6,\,22.0$~\AA)
were fitted at the same time, but the wavelength and strength were fixed across
all data sets because the absorber is considered to be independent of the
backlight spectrum. Our method therefore reduces any differences in the
continuum between data sets.

Next, absorption parameters for each line feature were further determined by
fitting only in a narrow spectral range centered on each feature. The
powerlaw indices of each  data set (determined by fitting over 
$21.3<\lambda<22.5$~\AA) were held fixed and only their normalizations were
allowed to vary. The fitting ranges were ($21.4<\lambda<21.8$~\AA),
($21.9<\lambda<22.1$~\AA) and ($22.2<\lambda<22.4$~\AA), respectively, for the
three detected features listed in Table~\ref{tab:mrk_detections_table}.

\begin{deluxetable}{cclccrrlr}
\scriptsize
\tablecolumns{9}
\tablewidth{0pc}
\tablecaption{Joint fitting results and line detection limits for
  datasets in Table~\ref{tab:mrk_dataset_table}\label{tab:mrk_detections_table}}
\tablehead{
  \colhead{Feature} &
  \colhead{$\lambda_{meas}$\tablenotemark{a}} & 
  \colhead{$\Delta\lambda$\tablenotemark{a}}&
  \colhead{ID\tablenotemark{b}}&
  \colhead{z}&
  \colhead{$\rm W_\lambda$\tablenotemark{a}} &
  \colhead{$\rm S/N$\tablenotemark{c}} &
  \colhead{$\rm Log\,N_i$\tablenotemark{d}} &
  \colhead{$\chi_\nu^2$(d.o.f.)}}
\startdata
1  &  21.595 & $^{+\phn 3.2}_{-\phn3.0}$&  \ion{O}{7} & $\equiv 0$    & $13.6^{+1.1}_{-1.1}$  &25.6(20.4)    & $15.763^{+0.031}_{-0.046}$ & 0.951(726) \\
2  &  22.022 & $^{+10.0}_{-10.0}$       &  \ion{O}{6} & $0.0004 \pm .0005$  & $3.3^{+1.5}_{-1.4}$& 6.3\phn(4.2) & $15.22^{\phn+0.16}_{\phn-0.23}$   & 0.902(332) \\
3  &  22.290 & $^{+21.1}_{-19.9}$       & (\ion{O}{7})&$0.0318\pm .0010$ & $2.1^{+1.4}_{-1.3}$& 4.0\phn(2.6) & $14.93^{\phn+0.43}_{\phn-0.22}$   & 1.190(347) \\
\cutinhead{Non--detections}
&
$\lambda_{input}$\tablenotemark{e} & 
\multicolumn{1}{c}{$\Delta\lambda$\tablenotemark{e}}&
ID\tablenotemark{b}&
z\tablenotemark{e}&
\multicolumn{1}{c}{$\rm W_\lambda$\tablenotemark{a}} &
&
\multicolumn{1}{c}{$\lceil\rm Log\,N_i\rceil$\tablenotemark{f}} &
\colhead{$\chi_\nu^2$(d.o.f.)}\\
\hline
\nodata & 21.85 & $^{+20}_{-20}$& (\ion{O}{7})&$0.011^{+.001}_{-.001}$ & $\phn0\phn^{+1.8}_{-1.4}$& \multicolumn{1}{c}{\nodata} & 14.78(15.01)  & 1.025(346) \\
\nodata & 22.20 & $^{+20}_{-20}$& (\ion{O}{7})&$0.028^{+.001}_{-.001}$ & $\phn0\phn^{+2.0}_{-1.2}$& \multicolumn{1}{c}{\nodata} & 14.81(15.06)  & 1.229(347) \\
\nodata & 22.32 & $^{+20}_{-20}$& (\ion{O}{7})&$0.033^{+.001}_{-.001}$ & $\phn0\phn^{+2.7}_{-0.5}$& \multicolumn{1}{c}{\nodata} & 14.93(14.93)  & 1.182(348) \\
\enddata
\tablenotetext{a}{
  Quoted uncertainties are for 90\% confidence limits ($\Delta\chi^2 = 2.706$). Wavelength uncertainties and equivalent widths are given in m\AA.}
\tablenotetext{b}{Parenthesized entries are tentative. \ion{O}{7} and \ion{O}{6} represent the K$\alpha$ blends $\lambda$21.602\AA\ and $\lambda$22.02\AA, respectively.}
\tablenotetext{c}{Significance of each detection. The first number given is $\sqrt{\Delta\chi^2}$ in the well--sampled continuum limit. Parenthesized values are for narrow spectral range fitting: $\Delta\lambda=$0.4\AA\ for Feature~1, $\Delta\lambda=$0.2\AA\ for other features.}
\tablenotetext{d}{Column densities corresponding to each feature. Feature~1 was fit best when using a turbulent 
  velocity parameter $\rm v=330\,km\,s^{-1}$; the other two lines were fit using a fixed $\rm v=100\,km\,s^{-1}$. 
  The effective oscillator strengths assumed for \ion{O}{7} and \ion{O}{6} were 0.695 and 0.525, respectively.}
\tablenotetext{e}{Quantities were taken from the detections of \citet{Nicastro05}.}
\tablenotetext{f}{90\% confidence upper limits to absorption column
  density. The first number corresponds to fitting the line feature at the input wavelength
  tablulated by \citet{Nicastro05}; the parenthesized value
  corresponds to allowing the absorber wavelength to vary freely
  within the 40~m\AA\ band centered on $\lambda_{input}$, the wavelength uncertainty they assumed for \chandra\ \letgs.}
\end{deluxetable}

To fit absorption line profiles and to estimate equivalent widths, we generated
response matrices with energy bin densities that oversample the RGS spectral
resolution by a factor of about 5 ($\Delta\lambda\sim\lambda/1800$), sufficient
to model narrow features in the spectra. For the intrinsic line
profiles we used the Voigt profile, evaluated appropriately for a
discrete model wavelength grid.

The fit parameters providing the best fits (for given temperature, atomic mass
and Doppler parameter) are given in the oscillator strength--column density
product ($fN_i$) and the equivalent width $W_\lambda$ for the line
(again, computed on the spectral model grid).

Three apparent absorption features stand out visually: at wavelengths $\lambda
\approx 21.60, 22.02, 22.29$~\AA. The results of formally fitting
these features are summarized  in the first three entries of
Table~\ref{tab:mrk_detections_table}.

The strongest line 
% at 21.595~\AA\ was reported previously \citep{Nicastro01,
% Rasmussen03a, Paerels03, Nicastro05, Williams05} and is identified
% as 
is identified with the 
the \ion{O}{7}~$w$ transition 
at $z=0$
(21.602~\AA). The best fit wavelength is off by
7~m\AA\ (or 100~km\,s$^{-1}$) from the expected value and formally outside of
the measured 90~\% confidence limit range, by 4~m\AA; however this mismatch is
smaller than the overall wavelength scale zero point uncertainty. Therefore we
assume that it is the 21.602~\AA\ line feature at $z=0$, and it
provides a local wavelength fiducial for the remaining features in the spectrum.

The next strongest line is coincident with the dominant inner shell excitation
doublet of \ion{O}{6}, also at $z=0$, which has a laboratory
wavelength of 
22.0194$\pm$0.0016~\AA\ \citep{Schmidt04}. This line is
also quite prominent in Figure~\ref{fig:mrk421_resids}.
The measured strength of this line ($W_\lambda \sim 3.3$~m\AA) is marginally
consistent with the \chandra\ \letgs\ measurement of $2.4 \pm 0.9$~m\AA\
\citep{Williams05} but the inferred column density raises some
concern, because it is large compared to the column density in
\ion{O}{6} resolved by FUSE \citep{Savage05}. The narrow feature
immediately to the right of the \ion{O}{6} inner shell feature appears
to be narrower than the instrumental resolution. The 
source of this fluctuation could be instrumental, but an exhaustive
search for its origin has not been performed.
Unfortunately the wavelength range is only covered by 
RGS1\footnote{See Appendix~\ref{ap2} for a study of typical
  systematics intrinsic to the data, performed by comparing
  differences in the recorded spectrum between the two \rgs\ models.} 
and an independent cross check on this feature is not
available.
% original
%  RGS1 due to a
%  failure in RGS2 and therefore an independent cross check on this
%  feature is not available.

Finally, the weakest, $\sim2\sigma$ feature close to 22.29~\AA\ is
suggestive of 
absorption by \ion{O}{7} 21.602~\AA\ in the host galaxy
($0.031<z<0.033$). The 90\% confidence range for the feature redshift
is marginally consistent with a currently accepted redshift of
$z=0.0308$ for \mrk\ \citep{Ulrich75}. They did not give an
accuracy estimate for their redshift but an inspection of their
published plot suggests it may be of the order of 0.001. 

\subsection{Weak line feature search}

\begin{figure}
\includegraphics[angle=-90,width=\columnwidth]{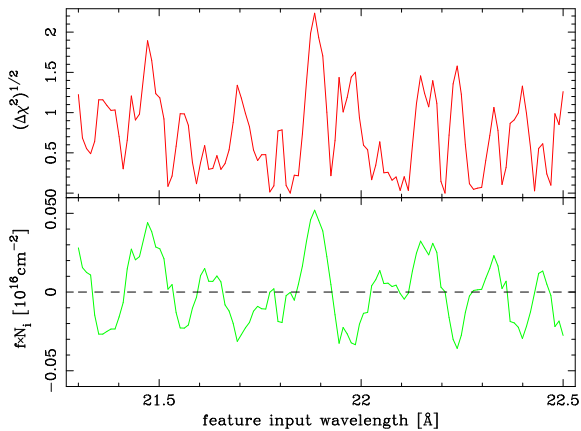}
\caption{Results of the search for narrow, unresolved features in the
  \ion{O}{7} region of the spectrum. The positive detections 
  identified in Table~\ref{tab:mrk_detections_table} have been
  included in the underlying model. The top plot shows
  $\sqrt{\Delta\chi^2}$ (in units of $\sigma$) between the best
  fit local continuum level and adding in a line--like feature at the
  center of the data band where the fitting
  was performed. The bottom plot gives the best fit parameter $fN_i$
  which encodes the polarity and equivalent width of the best fit
  feature. 
  \label{fig:featuresearch}}
\end{figure}

\begin{figure}
\centering
\includegraphics[angle=-90,width=\columnwidth]{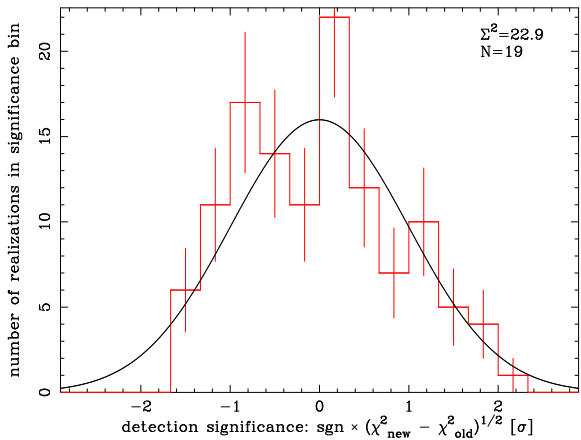}
\caption{The distribution of ``detections'' in the narrow feature
  search results ({\it cf.}
  Fig.~\ref{fig:featuresearch}). 
  The distribution appears to be
  consistent with what is expected, based on the $\chi^2$ and $F$
  distributions (the overplotted curve).
  There is no indication for additional intervening absorption
  systems at this effective sensitivity level ($\sigma$
  corresponds to $\delta W_\lambda \sim 0.95\,$m\AA). 
\label{fig:sig_dist}}
\end{figure}

To identify any other possible spectral features in the wavelength range of
interest, we have performed a search by fitting for lines within localized
spectral ranges. The fitting ranges were 4 resolution elements in width
(0.2~\AA) and were centered on the hypothetical feature wavelength. Allowing
only the normalizations of the local continua to vary and with the powerlaw
indices fixed, the narrow spectral range was fitted. Then a line
feature (either in absorption or emission) was introduced to the model and its best fit amplitude
was determined, along with the change in $\chi^2$.  This process was
performed on a fine, uniformly spaced grid of wavelengths within the
search region. Results of this process are given in
Fig.~\ref{fig:featuresearch}. The search is effectively a localized
test of the null hypothesis (no 
line), performed over the spectral range. While some features appear
to be fit with significances of 2$\sigma$, the overall distribution of
these ``detections'' is well behaved (Fig.~\ref{fig:sig_dist}) and the
number of 2$\sigma$ detections is not greater 
than expected, according to the $\chi^2$ and $F$ distributions. The
typical null hypothesis probability for obtaining our value for $F$ is
8.5\%. We conclude that at the working sensitivity level of the blind
search, the data are consistent with {\it no excess in features} above
the 2$\sigma$ level (1.9~m\AA; $N_{OVII} \sim 7\times
10^{14}$\,cm$^{-2}$) that may trace intervening WHIM toward \mrk. 

\section{Comparison to Published Results}

\subsection{Comparison to \citet{Williams05}}
We immediately notice that our measured $W_\lambda$ values are
significantly larger ($\gtrsim$30\%) than those tabulated based on the
\chandra\ results. We investigated reasons for this, and found that
the discrepancy is most likely due to the combined effect of 
several factors. First, it should be noted that $W_\lambda$
values derived from order-unsorted data ({\it LETG/HRC-S}) can depend
strongly on properties of the underlying continuum model and
the assumed column density. When we fit the \letgs\ data ourselves
\citep{Kaastra06}, we obtain $W_\lambda$ values for the order sorted
({\it LETG/ACIS-S}) observations that exceed the published 
\citep{Williams05} values by roughly 20\%. Moreover, when we attempt
to estimate $W_\lambda$ values graphically from their Figure~1(c)
we are able to reproduce their tabulated values precisely. We found
this surprising because the presence of spectral contamination in
their composite data set (estimated at about 20\%) should reduce the
apparent $W_\lambda$ value correspondingly. We therefore believe that 
contaminating contributions to the continuum level had not been 
properly subtracted prior to their $W_\lambda$ estimation, thereby
resulting in systematically low equivalent widths. Because
the discussion of their data analysis refers to \citet{Nicastro05} for
a thorough description, we infer that the systematic underestimation
affects both of these papers.

The presence of finite spectral contamination by scattering 
further decreases apparent equivalent widths. While integrated
contributions of this sort are considered small, we note
that off-diagonal terms are neglected in off--the--shelf
\chandra\ matrices:  Response values ({\it e.g.}, {\tt  
http://asc.harvard.edu\-/cal\-/Links\-/Letg\-/User\-/Hrc\_QE\-/ea\_index.html})
are identically zero for off--diagonal elements where
$|m\Delta\lambda|>75\rm\,m\AA$.

On the \rgs\ side, it is certainly conceivable that estimated
$W_\lambda$ values may be systematically high or low by a relatively
small ($\sim$10\%) amount. Pre-flight calibration activities included
illuminating individual \rgs\ gratings to measure their scatter
distribution over various angular scales. Interpretation of the raw
data naturally placed upper limits to true scatter contributions, but
improper accounting for of any spectral contamination in the source or
scatter in the beamline each affect the apparent scatter
amplitude off of the gratings. A modest overestimation in the grating
scatter amplitude in the \rgs\ physical model also lead to
overestimates in $W_\lambda$ measures, according to our method, by
approximately the same amount.  

It is evident, therefore, that the $z$=0 \ion{O}{7} feature in the
spectrum is reported inconsistently in units of m\AA.
Unless otherwise noted in the following discussion, we disregard this
fact and use tabulated entries for absorption line strengths
($W_\lambda$) taken at face value. 

\begin{figure}
\centering
\includegraphics[angle=-90,width=\columnwidth]{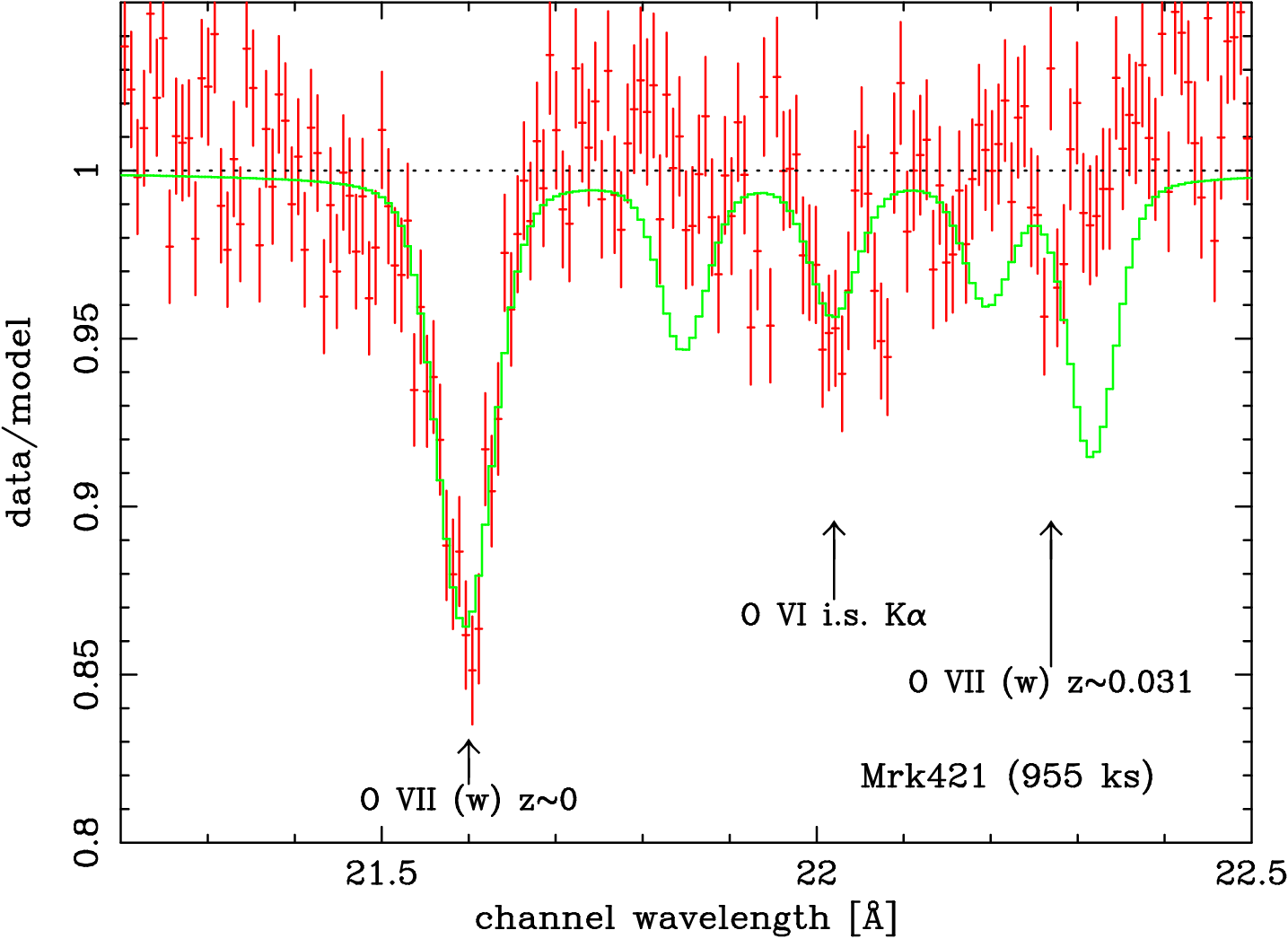}
\caption{A comparison of the 955~ks absorption spectrum toward \mrk\
  to the absorption line pattern of the \chandra\ \letgs\ spectrum
  \citep{Nicastro05,Williams05}.
  \label{fig:folded_spectrum}}
\end{figure}

\subsection{Comparison to \citet{Nicastro05}}
As we have already shown above, a blind search of the RGS spectrum for
narrow absorption lines in the $21.3-22.5$~\AA\ wavelength range
yields no evidence for the presence of lines that could indicate
absorption by intervening WHIM. Furthermore, directly fitting for the
absorption features identified in the \chandra\ \letgs\ spectrum
reported by \citet{Nicastro05} yielded only null results ({\it cf.}
the three non--detections tabulated in
Table~\ref{tab:mrk_detections_table}).  

We are left to compare the \xmm\ \rgs\ absorption spectrum toward
\mrk\ we obtained to the folded spectrum that is consistent with the
reported intervening absorbers. This is given in
Figure~\ref{fig:folded_spectrum}. To generate this comparison we took
the feature equivalent widths tabulated in the two \letgs\ papers
\citep{Nicastro05,Williams05} together with a quantitative estimate
for the $z=0.033$ feature which was not tabulated in either paper, to
produce a pattern of absorption features. The pattern 
amplitude was then scaled to obtain agreement between the data
and the folded model for the \ion{O}{7} $w$ ($z=0$)
feature as given in Table~\ref{tab:mrk_detections_table}.

The comparison between the \rgs\ data and the synthesized \letgs\ model
(folded through the \rgs\ response) is striking. Overall, two features are in
reasonable agreement, while three features are not. All three features
that are inconsistent 
are the purported intervening \ion{O}{7}~$w$ absorption systems at
redshifts $z\approx$0.011, 0.027 \& 0.033. Their presence, at the
contrast seen in the \letgs\ spectrum is ruled out at the 3.5~$\sigma$,
2.8~$\sigma$ and 5.2~$\sigma$ levels, respectively\footnote{These
  exclusions are based on $W_\lambda$ {\it face value} tabulations and
  not on the line strengths reflected in
  Fig.~\ref{fig:folded_spectrum}}. 
We can conclude that the features in the \chandra\ data as reported by
\citet{Nicastro05} are in disagreement with the results of our
analysis of the \rgs\ data.
% original
%  We can conclude that the features analyzed in the \chandra\ \letgs\
%  spectrum are either statistical fluctuations or were induced by
%  systematic errors in that instrument. 

\subsection{Comparison to \citet{Williams06}\label{comp_to_williams}}

\citet{Williams06} searched for the z$>0$ absorption lines
in a subset of RGS data. 
% original
%  However, they described the quality of their
%  accumulated data disappointing, and suffered impediments to performing
%  detailed quantitative analysis on data sets that were otherwise very
%  rich in counting statistics. 
They attributed their inability to confirm the intervening WHIM
features discovered by \citet{Nicastro05} to a host of instrumental
problems intrinsic to the \rgs, and declared their non-detection fully
consistent with the \chandra\ measurement. 
We have demonstrated that a careful analysis is not in agreement with
this conclusion. 
We have successfully used
the \xmm\ \rgs\ to produce and analyze data sets that are minimally
affected by instrumental systematics.  Our strong exclusion of 
{\it only} the redshifted features detected by \chandra\ 
makes the claim for the discovery of the WHIM 
dubious.
% original
%  highly suspect.

By studying the analysis approach described in \citet{Williams06}, we
have ascertained 
that they
% original
%  where they went wrong. 
%  In particular, they 
introduced certain problems in the method for co-adding
the 14
observations\footnote{http://www.astronomy.ohio-state.edu/\-$\sim$smita/\-xmmrsp/}.
For a source as variable as \mrk, where the spectra are accumulated 
from a large number of pointings characterized by small offsets,
certain systematics will naturally be introduced into the data
wherever bad detector regions exist\footnote{See the fourth item above
  in our list of systematics sources}.  The key lies in understanding
how non--continuous detector coverage can be used to still yield
results that are not riddled with artifacts from that
coverage. Because the countrates in the \rgs\ 
% original
%  have in fact 
varied by as much as a factor of 4 for the observations they used,
a single bad column that was properly ignored 
in the data pipeline 
(but improperly compensated for in modeling)
can conceivably introduce multi-modal features with
equivalent widths as large as 
6~m\AA\footnote{
  For two observations co--added, induced features (due to ignored
  columns) should have equivalent widths of order $W_\lambda^{sys}
  \lesssim \left({1+1/\tau \over
    1+\rho}-1\right)\times\Delta\lambda_c$, where $\tau$, $\rho$, and
  $\Delta\lambda_c$ are the integration time ratio, countrate ratio
  and spectral width of a channel, respectively}.   
This is a factor of 10 or so greater than the 1~$\sigma$ counting
statistics limit. 

While we accumulated counting statistics to a different level (26000
{\it vs.} their 12500 cts per 50~m\AA), this fact alone should have
had an inconsequential impact on the overall sensitivity to spectral
features of the significance detected by \chandra\ \letgs. 
We suggest that \citet{Williams06} did not exercise proper care in
collecting up large quantities of data for the purpose of measuring
faint spectral features. 
% original
%  This was undoubtedly the primary objective of their analysis. 

Irrespective of our mutually opposing conclusions, secondary reasons
for obtaining different quality spectra using the \rgs\ exist, 
and we list them here: 

\begin{enumerate}
\item 
  Our data were not added together, but fitting residuals were
  averaged for display purposes. Each dataset was analyzed with its
  corresponding response matrix, as described above. This naturally
  led to nearly complete spectral coverage over the range of interest.
\item 
  Large countrate fluctuations ($\sim$5\%) from channel to channel
  are seen in their spectrum (their Fig.~1), and are suggestive of
  aliasing problems. Similar fluctuations are sometimes seen only
  in the longest wavelength range of the \rgs\ and are a result of an
  overzealous hot pixel finder. The problem can be ameliorated
  significantly by altering the extraction region definitions or 
  parameters of the hot pixel finder. The problem is normally not
  noticeable in the shorter wavelength ranges ($\lambda\lesssim
  29$\AA). The fluctuations seen in all wavelength ranges of their
  spectrum are therefore suggestive of a different problem. Instead,
  they are probably due to the co--adding approach they took, which we
  referred to above.
\item 
  They state that pointing offsets exceeding 15\arcsec\ were not
  included because spectral resolution is degraded there. This an
  unnecessary restriction that led to smaller counting statistics for
  them. For example, we have combined data from multiple offset 
  pointings, and rely on the aspect correction algorithms for the 
  \rgs\ to align spectral data. The \rgs\ spectral resolution does not
% original
%  in fact 
  really
  degrade significantly even with $\sim$few arc\-minute scale
  pointing offsets. 
\item
  They claim that the presence of a `detector feature' rendered the
  $z\sim$0.011 feature unmeasurable in the RGS. The SAS provides means
  to exclude specific detector regions, both from the data and from
  the response. We have shown that with proper care, systematics
  arising from detector regions may be almost completely 
  removed. The machinery that provides this capability is available
  and is central to the \rgs\ branch of the SAS.
% original
%  widely available, and we suggest that users be aware of it.
\end{enumerate}

We summarize our disagreement with \citet{Williams06} as follows:
While the systematics introduced by their method hindered
spectral interpretation, those problems are 
not insurmountable. Signatures of systematics can be
distinguished from instrumental PSFs, and the equivalent widths of the
features reported by \citet{Nicastro05} are large enough to see
by overplotting the data ({\it cf.} Fig.~\ref{fig:folded_spectrum}).
% original
%  Thus 
As we have demonstrated above,
the \rgs\ data can indeed be used to confirm or deny presence of the 
weak spectral features that constitute the 
\chandra\ \letgs\ \citep{Nicastro05} detections. 

\section{Conclusions}

We have analyzed a very deep spectrum of \mrk\ obtained with the \rgs\ 
on \xmm. In 1~Ms of exposure, we collected over 26000 counts per
50~m\AA\ resolution element,which gives us high sensitivity to the
detection of narrow absorption lines. We describe the detailed
procedure by which we reduced the data, which results in a 
continuum spectrum with noise properties that are dominated by
statistical fluctuations.
The deep continuum spectrum of \mrk\ is well enough understood that it allows 
us to detect real absorption lines of equivalent width $>2.4$~m\AA\
with 99\% confidence. 

Focusing attention on the $21.3-22.5$~\AA\ band, which contains the
wavelength range in which intervening \ion{O}{7}~$w$ absorption lines
along the line of sight to \mrk\ should be contained, we find no evidence for redshifted absorption lines. We place a 
2~$\sigma$ upper limit of 1.9~m\AA\ on the equivalent with of any
narrow absorption line over the quoted wavelength band. 

This finding is in clear conflict with the 
claim of the detection of
narrow absorption lines at 21.85 and 22.20 \AA\ in the deep \chandra\
\letgs\ spectrum of the source, which would naturally correspond to
\ion{O}{7}~$w$ lines redshifted to $z = 0.011$ and $z = 0.027$,
respectively. We find that absorption lines at these positions, at the
equivalent widths claimed on the basis of the \chandra\ \letgs\
spectrum, are excluded with high confidence (3.5 and
2.8~$\sigma$, respectively). The exclusion becomes only stronger
when we use the mutually detected \ion{O}{7} $z=0$ feature as
an absorption fiducial to align the $W_\lambda$ values measured across
instruments as represented in Figure~\ref{fig:folded_spectrum} (5.3
and 3.8~$\sigma$, respectively). 

We conclude that the detection of absorption by intergalactic He-like
oxygen in filaments of column densities $N_i \approx 8.5\times
10^{14}$~cm$^{-2}$ is all but excluded (with 99\% confidence) by our
data. 
% original
%  The simplest interpretation of the discrepancy between the \xmm\
%  and \chandra\ spectroscopic results is that the apparent positive
%  detection in the \chandra\ data is a statistical
%  fluctuation. For further comments on this issue we refer to a 
%  companion paper \citep{Kaastra06}.
Possible explanations for the apparent discrepancy between the \xmm\
and \chandra\ spectroscopic results are investigated in an
accompanying paper \citep{Kaastra06}.

The non-detection of intervening absorption, even in the deepest
exposures that can currently be contemplated, on the brightest suitable
extragalactic continuum source, is probably not surprising. The
redshift of \mrk\ is relatively small, and the {\it a priori}
probability, as predicted by recent cosmological gas dynamics
simulations, of having an intervening filament with a column density
that is readily and convincingly detectable with the current 
instrumentation, is relatively small: 
for $Z\sim 0.1\, Z_\sun$, ${dP \over dz}\times z_{Mrk421} \sim 0.05 -
0.2$ for Log$(N_i)>15.0$ and $14.6$, respectively
\citep{FBC02,Chen03}. The most 
uncertain parameter in these calculations is the absolute
oxygen abundance, which can currently not be calculated with
confidence from first principles.  The null detections therefore do
not yet constrain these calculations in a meaningful way. The
detection of the `missing baryons' remains a challenge, best addressed 
with higher resolution, higher sensitivity spectrometers.

\begin{acknowledgements}
We acknowledge several discussions with Fabrizio Nicastro in which
we shared our data and analysis, following his WHIM discovery claims. 
% We thank the Editor-in-Chief at ApJ for strongly suggesting that we
% make our manuscript to be more collegial in tone.
This work was supported by NASA.
SRON is supported financially by NWO, the Netherlands Organization for
Scientific Research.
\end{acknowledgements}

\bibliographystyle{apj}

\begin{thebibliography}{31}
\expandafter\ifx\csname natexlab\endcsname\relax\def\natexlab#1{#1}\fi

\bibitem[{{Arnaud}(1996)}]{Arnaud96}
{Arnaud}, K.~A. 1996, in ASP Conf. Ser. 101: Astronomical Data Analysis
  Software and Systems V, ed. G.~H. {Jacoby} \& J.~{Barnes}, 17--+

\bibitem[{{Barcons} {et~al.}(2005){Barcons}, {Paerels}, {Carrera}, {Ceballos},
  \& {Sako}}]{Barcons05}
{Barcons}, X., {Paerels}, F.~B.~S., {Carrera}, F.~J., {Ceballos}, M.~T., \&
  {Sako}, M. 2005, \mnras, 359, 1549

\bibitem[{{Brinkman} {et~al.}(2000){Brinkman}, {Gunsing}, {Kaastra}, {van der
  Meer}, {Mewe}, {Paerels}, {Raassen}, {van Rooijen}, {Braeuninger}, {Burwitz},
  {Hartner}, {Kettenring}, {Predehl}, {Drake}, {Johnson}, {Kenter}, {Kraft},
  {Murray}, {Ratzlaff}, \& {Wargelin}}]{Brinkman00}
{Brinkman}, B.~C., {Gunsing}, T., {Kaastra}, J.~S., {van der Meer}, R., {Mewe},
  R., {Paerels}, F.~B., {Raassen}, T., {van Rooijen}, J., {Braeuninger}, H.~W.,
  {Burwitz}, V., {Hartner}, G.~D., {Kettenring}, G., {Predehl}, P., {Drake},
  J.~J., {Johnson}, C.~O., {Kenter}, A.~T., {Kraft}, R.~P., {Murray}, S.~S.,
  {Ratzlaff}, P.~W., \& {Wargelin}, B.~J. 2000, in Proc. SPIE Vol. 4012, p.
  81-90, X-Ray Optics, Instruments, and Missions III, Joachim E. Truemper;
  Bernd Aschenbach; Eds., ed. J.~E. {Truemper} \& B.~{Aschenbach}, 81--90

\bibitem[{{Burles} \& {Tytler}(1997)}]{Burles97}
{Burles}, S., \& {Tytler}, D. 1997, \aj, 114, 1330

\bibitem[{{Burles} \& {Tytler}(1998)}]{Burles98}
---. 1998, \apj, 507, 732

\bibitem[{{Cagnoni} {et~al.}(2004){Cagnoni}, {Nicastro}, {Maraschi}, {Treves},
  \& {Tavecchio}}]{Cagnoni04}
{Cagnoni}, I., {Nicastro}, F., {Maraschi}, L., {Treves}, A., \& {Tavecchio}, F.
  2004, \apj, 603, 449

\bibitem[{{Canizares} {et~al.}(2000){Canizares}, {Huenemoerder}, {Davis},
  {Dewey}, {Flanagan}, {Houck}, {Markert}, {Marshall}, {Schattenburg},
  {Schulz}, {Wise}, {Drake}, \& {Brickhouse}}]{Canizares00}
{Canizares}, C.~R., {Huenemoerder}, D.~P., {Davis}, D.~S., {Dewey}, D.,
  {Flanagan}, K.~A., {Houck}, J., {Markert}, T.~H., {Marshall}, H.~L.,
  {Schattenburg}, M.~L., {Schulz}, N.~S., {Wise}, M., {Drake}, J.~J., \&
  {Brickhouse}, N.~S. 2000, \apjl, 539, L41

\bibitem[{{Cen} \& {Ostriker}(1999)}]{Cen99}
{Cen}, R., \& {Ostriker}, J.~P. 1999, \apj, 514, 1

\bibitem[{{Chen} {et~al.}(2003){Chen}, {Weinberg}, {Katz}, \&
  {Dav{\'e}}}]{Chen03}
{Chen}, X., {Weinberg}, D.~H., {Katz}, N., \& {Dav{\'e}}, R. 2003, \apj, 594,
  42

\bibitem[{{Cowie} {et~al.}(1995){Cowie}, {Songaila}, {Kim}, \& {Hu}}]{Cowie95}
{Cowie}, L.~L., {Songaila}, A., {Kim}, T.-S., \& {Hu}, E.~M. 1995, \aj, 109,
  1522

\bibitem[{{Croft} {et~al.}(2001){Croft}, {Di Matteo}, {Dav{\'e}}, {Hernquist},
  {Katz}, {Fardal}, \& {Weinberg}}]{Croft01}
{Croft}, R.~A.~C., {Di Matteo}, T., {Dav{\'e}}, R., {Hernquist}, L., {Katz},
  N., {Fardal}, M.~A., \& {Weinberg}, D.~H. 2001, \apj, 557, 67

\bibitem[{{den Herder} {et~al.}(2001){den Herder}, {Brinkman}, {Kahn},
  {Branduardi-Raymont}, {Thomsen}, {Aarts}, {Audard}, {Bixler}, {den Boggende},
  {Cottam}, {Decker}, {Dubbeldam}, {Erd}, {Goulooze}, {G{\"u}del}, {Guttridge},
  {Hailey}, {Janabi}, {Kaastra}, {de Korte}, {van Leeuwen}, {Mauche},
  {McCalden}, {Mewe}, {Naber}, {Paerels}, {Peterson}, {Rasmussen}, {Rees},
  {Sakelliou}, {Sako}, {Spodek}, {Stern}, {Tamura}, {Tandy}, {de Vries},
  {Welch}, \& {Zehnder}}]{denHerder01}
{den Herder}, J.~W., {Brinkman}, A.~C., {Kahn}, S.~M., {Branduardi-Raymont},
  G., {Thomsen}, K., {Aarts}, H., {Audard}, M., {Bixler}, J.~V., {den
  Boggende}, A.~J., {Cottam}, J., {Decker}, T., {Dubbeldam}, L., {Erd}, C.,
  {Goulooze}, H., {G{\"u}del}, M., {Guttridge}, P., {Hailey}, C.~J., {Janabi},
  K.~A., {Kaastra}, J.~S., {de Korte}, P.~A.~J., {van Leeuwen}, B.~J.,
  {Mauche}, C., {McCalden}, A.~J., {Mewe}, R., {Naber}, A., {Paerels}, F.~B.,
  {Peterson}, J.~R., {Rasmussen}, A.~P., {Rees}, K., {Sakelliou}, I., {Sako},
  M., {Spodek}, J., {Stern}, M., {Tamura}, T., {Tandy}, J., {de Vries}, C.~P.,
  {Welch}, S., \& {Zehnder}, A. 2001, \aap, 365, L7

\bibitem[{{Fang} {et~al.}(2002{\natexlab{a}}){Fang}, {Bryan}, \&
  {Canizares}}]{FBC02}
{Fang}, T., {Bryan}, G.~L., \& {Canizares}, C.~R. 2002{\natexlab{a}}, \apj,
  564, 604

\bibitem[{{Fang} \& {Canizares}(2000)}]{Fang00}
{Fang}, T., \& {Canizares}, C.~R. 2000, \apj, 539, 532

\bibitem[{{Fang} {et~al.}(2001){Fang}, {Marshall}, {Bryan}, \&
  {Canizares}}]{Fang01}
{Fang}, T., {Marshall}, H.~L., {Bryan}, G.~L., \& {Canizares}, C.~R. 2001,
  \apj, 555, 356

\bibitem[{{Fang} {et~al.}(2002{\natexlab{b}}){Fang}, {Marshall}, {Lee},
  {Davis}, \& {Canizares}}]{Fang02}
{Fang}, T., {Marshall}, H.~L., {Lee}, J.~C., {Davis}, D.~S., \& {Canizares},
  C.~R. 2002{\natexlab{b}}, \apjl, 572, L127

\bibitem[{{Kaastra} {et~al.}(2006){Kaastra}, {Werner}, {den~Herder}, {Paerels},
  {de~Plaa}, \& {Rasmussen}}]{Kaastra06}
{Kaastra}, J.~S., {Werner}, N., {den~Herder}, J.~W.~A., {Paerels}, F.~B.~S.,
  {de~Plaa}, J., \& {Rasmussen}, A.~P.~{de~Vries}, C.~P. 2006, \apj,
  (submitted)

\bibitem[{{Mathur} {et~al.}(2003){Mathur}, {Weinberg}, \& {Chen}}]{Mathur03}
{Mathur}, S., {Weinberg}, D.~H., \& {Chen}, X. 2003, \apj, 582, 82

\bibitem[{{Nicastro} {et~al.}(2001){Nicastro}, {Fruscione}, {Elvis},
  {Siemiginowska}, {Fiore}, \& {Bianchi}}]{Nicastro01}
{Nicastro}, F., {Fruscione}, A., {Elvis}, M., {Siemiginowska}, A., {Fiore}, F.,
  \& {Bianchi}, S. 2001, in ASP Conf. Ser. 234: X-ray Astronomy 2000, ed.
  R.~{Giacconi}, S.~{Serio}, \& L.~{Stella}, 511--+

\bibitem[{{Nicastro} {et~al.}(2005{\natexlab{a}}){Nicastro}, {Mathur}, {Elvis},
  {Drake}, {Fang}, {Fruscione}, {Krongold}, {Marshall}, {Williams}, \&
  {Zezas}}]{Nicastro05N}
{Nicastro}, F., {Mathur}, S., {Elvis}, M., {Drake}, J., {Fang}, T.,
  {Fruscione}, A., {Krongold}, Y., {Marshall}, H., {Williams}, R., \& {Zezas},
  A. 2005{\natexlab{a}}, \nat, 433, 495

\bibitem[{{Nicastro} {et~al.}(2005{\natexlab{b}}){Nicastro}, {Mathur}, {Elvis},
  {Drake}, {Fiore}, {Fang}, {Fruscione}, {Krongold}, {Marshall}, \&
  {Williams}}]{Nicastro05}
{Nicastro}, F., {Mathur}, S., {Elvis}, M., {Drake}, J., {Fiore}, F., {Fang},
  T., {Fruscione}, A., {Krongold}, Y., {Marshall}, H., \& {Williams}, R.
  2005{\natexlab{b}}, \apj, 629, 700

\bibitem[{{Nicastro} {et~al.}(2002){Nicastro}, {Zezas}, {Drake}, {Elvis},
  {Fiore}, {Fruscione}, {Marengo}, {Mathur}, \& {Bianchi}}]{Nicastro02}
{Nicastro}, F., {Zezas}, A., {Drake}, J., {Elvis}, M., {Fiore}, F.,
  {Fruscione}, A., {Marengo}, M., {Mathur}, S., \& {Bianchi}, S. 2002, \apj,
  573, 157

\bibitem[{{Paerels} {et~al.}(2003){Paerels}, {Rasmussen}, {Kahn}, {Herder}, \&
  {Vries}}]{Paerels03}
{Paerels}, F., {Rasmussen}, A., {Kahn}, S., {Herder}, J.~W., \& {Vries}, C.
  2003, {X-ray Absorption and Emission Spectroscopy of the Intergalactic Medium
  at Small Redshift}

\bibitem[{{Rasmussen} {et~al.}(2003{\natexlab{a}}){Rasmussen}, {Kahn}, \&
  {Paerels}}]{Rasmussen03a}
{Rasmussen}, A., {Kahn}, S.~M., \& {Paerels}, F. 2003{\natexlab{a}}, in ASSL
  Vol. 281: The IGM/Galaxy Connection. The Distribution of Baryons at z=0, ed.
  J.~L. {Rosenberg} \& M.~E. {Putman}, 109--+

\bibitem[{{Rasmussen} {et~al.}(2003{\natexlab{b}}){Rasmussen}, {Kahn},
  {Paerels}, {den Herder}, \& {de Vries}}]{Rasmussen03b}
{Rasmussen}, A., {Kahn}, S.~M., {Paerels}, F., {den Herder}, J., \& {de Vries},
  C. 2003{\natexlab{b}}, AAS/High Energy Astrophysics Division, 7, \#201

\bibitem[{{Ravasio} {et~al.}(2005){Ravasio}, {Tagliaferri}, {Pollock},
  {Ghisellini}, \& {Tavecchio}}]{Ravasio05}
{Ravasio}, M., {Tagliaferri}, G., {Pollock}, A.~M.~T., {Ghisellini}, G., \&
  {Tavecchio}, F. 2005, \aap, 438, 481

\bibitem[{{Savage} {et~al.}(2005){Savage}, {Wakker}, {Fox}, \&
  {Sembach}}]{Savage05}
{Savage}, B.~D., {Wakker}, B.~P., {Fox}, A.~J., \& {Sembach}, K.~R. 2005, \apj,
  619, 863

\bibitem[{{Schmidt} {et~al.}(2004){Schmidt}, {Beiersdorfer}, {Chen}, {Thorn},
  {Tr{\"a}bert}, \& {Behar}}]{Schmidt04}
{Schmidt}, M., {Beiersdorfer}, P., {Chen}, H., {Thorn}, D.~B., {Tr{\"a}bert},
  E., \& {Behar}, E. 2004, \apj, 604, 562

\bibitem[{{Ulrich} {et~al.}(1975){Ulrich}, {Kinman}, {Lynds}, {Rieke}, \&
  {Ekers}}]{Ulrich75}
{Ulrich}, M.-H., {Kinman}, T.~D., {Lynds}, C.~R., {Rieke}, G.~H., \& {Ekers},
  R.~D. 1975, \apj, 198, 261

\bibitem[{{Williams} {et~al.}(2006){Williams}, {Mathur}, {Nicastro}, \&
  {Elvis}}]{Williams06}
{Williams}, R.~J., {Mathur}, S., {Nicastro}, F., \& {Elvis}, M. 2006, \apjl,
  accepted for publication

\bibitem[{{Williams} {et~al.}(2005){Williams}, {Mathur}, {Nicastro}, {Elvis},
  {Drake}, {Fang}, {Fiore}, {Krongold}, {Wang}, \& {Yao}}]{Williams05}
{Williams}, R.~J., {Mathur}, S., {Nicastro}, F., {Elvis}, M., {Drake}, J.~J.,
  {Fang}, T., {Fiore}, F., {Krongold}, Y., {Wang}, Q.~D., \& {Yao}, Y. 2005,
  \apj, 631, 856

\end{thebibliography}

\appendix

\section{Effects of removing detector regions}
\label{ap1}
\begin{figure}
\includegraphics[angle=-90,width=0.9\columnwidth]{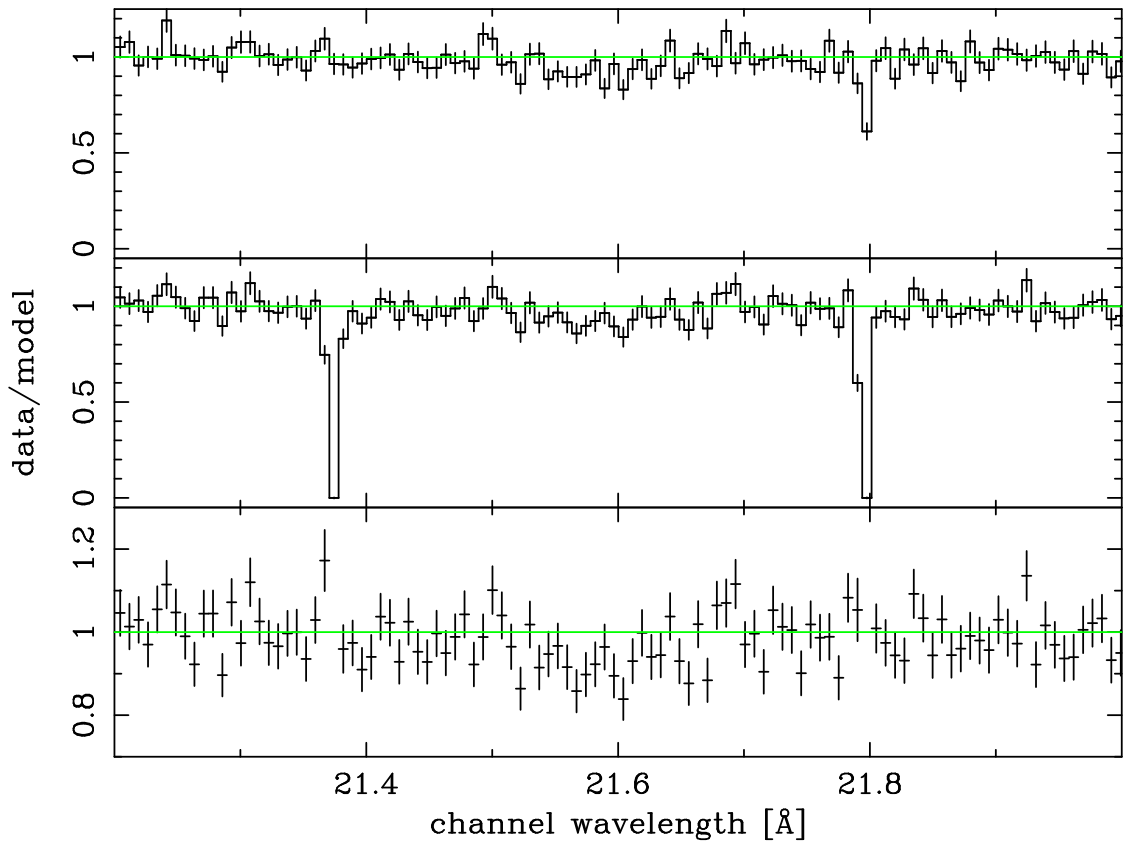}
\caption{\label{fig:cleaning_spectra} A demonstration of removing
detector regions in the data reduction process. A column with
non-standard response (close to 21.8\AA) is rejected, along with
a typical, randomly selected column (close to 21.37\AA). The three
spectra depict the same data at different iterations of the reduction.}
\end{figure}

A demonstration of the iterative analysis approach is given here.
The data in Figure~\ref{fig:cleaning_spectra} is from the dataset 0720,
and is plotted in terms of its ratio to a folded model. The top
histogram is the data processed with automatic detection and rejection
of flickering pixels, divided through with a continuum model folded
through its corresponding response matrix. An instrumental, sharp
feature located close to 21.8\AA, is identified with equivalent width
roughly 2.5~m\AA. Inspection of the same data binned by hardware
address reveals a single column with non--standard response. Because
the RGS normally operates in 3$\times$3 on--chip binning (OCB) mode,
the charge healthy and unhealthy columns alike are added prior to
digitization. In this case, it appears that the charge from a single
unhealthy column is added to the charge of two healthy ones prior to signal readout. The middle panel shows 
the same data after reprocessing, where the channel in question was
rejected. As a test of this method, we have also manually rejected an
additional randomly selected CCD column whose absense is seen at a
channel wavelength of $\sim$21.37\AA. These data are divided by the same
folded model and response matrix as in the top panel. The rejected
column affects more than one spectral channel because of the chosen
bin size. The bottom histogram finally shows the same data as in the
middle, but here is divided through by the folded model using the
updated response matrix, which include effects of the rejected
columns. While the column rejection process appears to introduce some
minor artifacts, substantial reduction in incidence of non-statistical
outliers is seen.  

\section{Weak spectral features arising from systematics}
\label{ap2}
We have shown above that with careful data analysis techniques, including
identification of problematic detector locations, systematics can be reduced to
a level which is comparable to the statistical limit. As statistics
build up in very deep, large signal--to--noise observations, beating down
systematics is critical in order for a proper identification of weak features.

The section of the spectrum discussed in this paper is recorded on a single CCD
in RGS1 which, due to a CCD failure early in the mission, has no counterpart
in RGS2. The quality of this CCD will be comparable to other CCD's on the RGS so 
the comparison between RGS1 and RGS2 can give us a
model independent estimate of the magnitude of systematic effects which may be present  
in RGS spectra in general and in the section of the spectrum analyzed in this paper in particular.

\begin{figure}
\centering
\includegraphics[angle=-90,width=0.9\columnwidth]{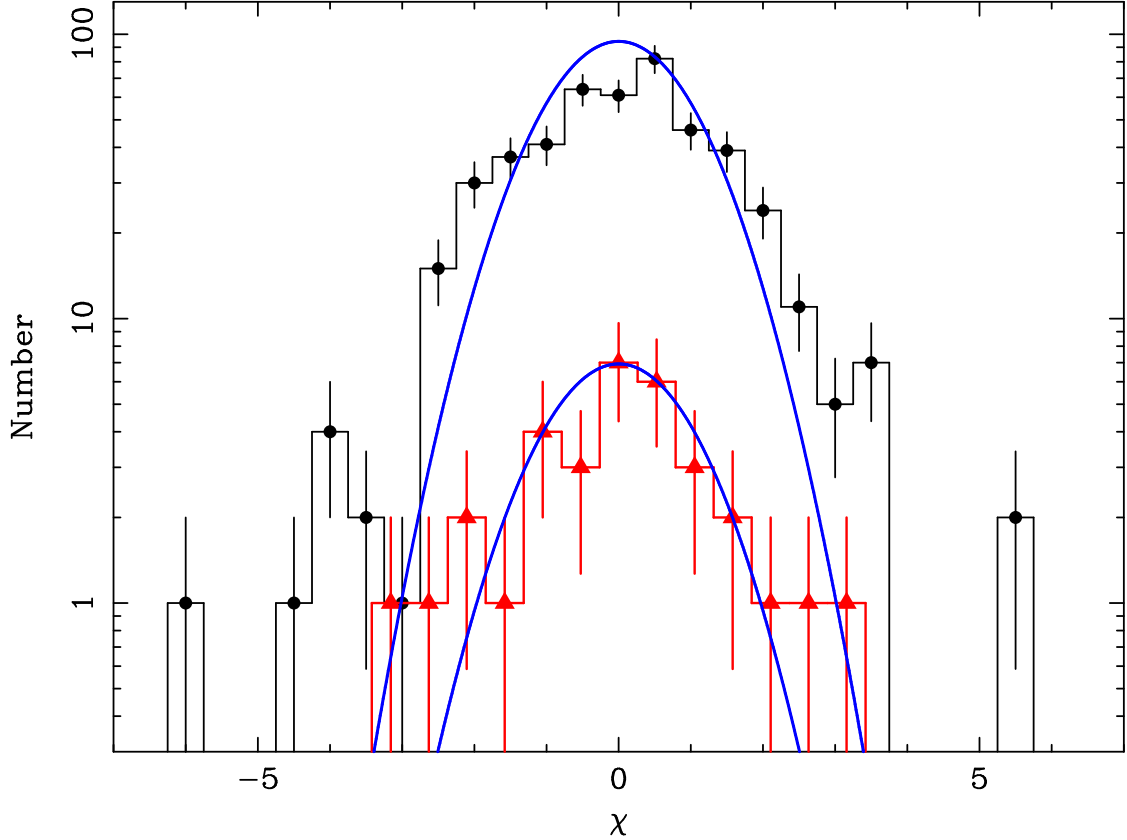}
\caption{The difference between RGS1 and RGS2, based on background
subtracted counts. The (top) histogram (filled circles) shows count
differences for single spectral bins between the two \rgs\ 
spectrometers, divided by the computed statistical variance of the
difference. The non--Gaussian tail component is assumed to be the
non--statistical, systematic contribution. The overplotted continuous line,
for comparison purpose, shows a strict Gaussian distribution, based on
count statistics only. The lower distribution (filled triangles) show
the distribution of differences between the model continuum and the
actual data for the analyzed section of RGS1, CCD4; the CCD which is
used to obtain the spectrum section discussed in this paper. Since
this spectrum contains a spectral range of only $\sim$1.2\AA, compared
to the overlapping coverage between RGS1 and RGS2 ($\sim$21\AA),
there is a factor of about 15 difference in number of data points
binned.  
The actual distributions appear indeed to be shifted with respect to
each other by a constant factor of about 10, confirming that our
estimate of systematic errors can also be applied to the section of
the spectrum discussed here.
\label{fig:hcmp}} 
\end{figure}

To evaluate remaining systematic errors in RGS  we have performed a model
independent analysis of the source spectrum seen by the twin RGS instruments. By
comparing the count ratio seen by the two instruments on a channel--by--channel
basis in wavelength coordinates and comparing this to an expected statistical distribution, the
non--statistical (or systematic) contribution may be resolved from the
statistical population as a wing. Figure~\ref{fig:hcmp} displays an example of
this distribution, with a Gaussian distribution (based on the channel counts only)
overlaid to guide the eye and to help visualize the non--statistical
contribution. 

Table~\ref{tab:stat_features} provides a tabulation of these quantities for bins
of 40 m\AA~ (the approximate width of the RGS PSF) per CCD pair. Based on these
results {\it if one only assumes purely statistical errors, there is about a 15\% chance
of falsely identifying a feature. For $>2\sigma$ features this corresponds to 3~m\AA~
absorption features.} It is stressed that the type of systematics discussed here can only 
result in absorption features and not in emission features, since all these effects can
only lower the effective area and not increase it.

Among the main causes for the non-statistical distribution in
Figure~\ref{fig:hcmp} will be 
the effects of small differences in CTE between columns in relation to the pulse-height data
selection window. Columns with clearly bad CTE will be manually discarded, but
small CTE differences may not be recognized. They will however, influence the
number of recorded photons due to the pulse-height selection window and hence will
result in non-statistical count rate differences. 

Recognition of individual systematic features, which are linked to fixed
detector positions, can only be done with high confidence if detector
coordinates change over time with respect to the wavelength
coordinates i.e. an effective dithering of the observations.  
In addition, systematic features too weak to be recognized will be
smeared-out by dithering, effectively decreasing their impact. It
would therefore be advisable to purposely dither long RGS
observations when detection of weak spectral features is desired.

\begin{table}
\centering
\caption{Consistency between RGS1 and RGS2 showing only those CCD's
which are operational 
in both RGSs. The flux and noise columns indicate the flux and
statistical noise in the \mrk\ spectrum for the specified CCD. 
$W_\lambda$ gives the equivalent width of a 1 $\sigma$ feature. 
The last column indicates what percentage of the data lies outside the
statistical 
Gaussian distribution.  
These results indicate that there is an overall probability of about
15\% for a $2 \sigma$ feature with a typical EW of  3~m\AA, when the
data is sampled for 40~m\AA\ bins. 
\label{tab:stat_features}   }
\begin{tabular}{c c c c r}
\hline
CCD &
Flux & 
Noise &
$W_\lambda$ &
Outside \\
 &
s$^{-1}$\,cm$^{-2}$ & 
s$^{-1}$\,cm$^{-2}$ & 
m\AA\ &
\% \\
 \hline
  1 & 0.0100 &  0.0004 &  1.55 &  5  \\
  2 & 0.0105 &  0.0004 &  1.48 &  12 \\
  3 & 0.0122 &  0.0005 &  1.59 &  14 \\
  - &	&	&	&	\\
  5 & 0.0140 &  0.0003 &  0.83 &  12 \\
  6 & 0.0145 &  0.0003 &  0.80 &  20 \\
  - & 	&	&	&	\\
  8 & 0.0140 &  0.0004 &  1.11 &  25 \\
  9 & 0.0130 &  0.0007 &  2.09 &  20 \\
\hline
\end{tabular}
\end{table}

\section{SAS data reduction prescriptions}
\label{ap3}
Almost all XMM observers use the standard XMM software (SAS) for their
analysis. We describe in this section how we
processed the \mrk\ observations using the SAS and dealt with the
systematic effects present in the data.

The SAS does handle some aspects of CCD systematics by
default, but some others require manual control. This is only relevant in the
search of weak features in spectra with a rather long observations. 
We used the  SAS version 6.5, but with
task `rgsenergy' version 2.0.3 (standard SAS version 6.5 uses `rgsenergy'
version 2.0.2). This allows CCD-pixel dependent offsets to be subtracted
(option: withdiagoffset=yes). These offsets are retrieved from diagnostics CCD
images, sampled at regular intervals during observations. In addition, the
default filter in the SAS to recognize and delete hot pixels was modified to
delete the hot pixels only, and not their neighbors (option:
rejflags="BAD\_SHAPE ON\_BADPIX ON\_WINDOW\_BORDER BELOW\_ACCEPTANCE" ).

The SAS removes detectable, high duty cycle (``hot'') pixels by
default. To identify additional bad pixels and 
columns {\it not} recognized by the SAS, the data are displayed in the CCD
coordinate reference frame as images of photon energy versus the CCD dispersion
axis. Since the individual \mrk\ observations have
slightly different pointings, real spectral features are spread out over a range
of CCD hardware coordinates, while CCD defects are  localized in these
coordinates. Such displays were used to  identify local CCD
defects. Using software 
written for this purpose, suspicious columns and pixels are found manually and
included in the SAS current calibration files (CCF) to be removed during the
subsequent processing. As described in the text, many of the columns identified in this way seem to suffer
from loss of charge transfer presumably due to local CCD radiation damage. In
the Energy versus CCD-column plot, X-ray events appear to be recorded at
much lower energies than for the neighboring columns - resulting in a
different detection efficiency into the local pulseheight window. Such
regions can currently only be identified manually, with the aid of
high counting statistics observations, such as those used here.
After modifying the CCF to include these bad columns, all data are processed
again through the SAS. 

The spectra generated for each individual observation were run through the task
{\tt rgsfluxer} to obtain individually fluxed spectra. Due to the source
variability these spectra were not simply added. Instead the spectra are all
normalized over a short wavelength interval and these normalized spectra are
added. The error is calculated taking into account the normalization.

\begin{figure}
\includegraphics[angle=-90,width=0.9\columnwidth]{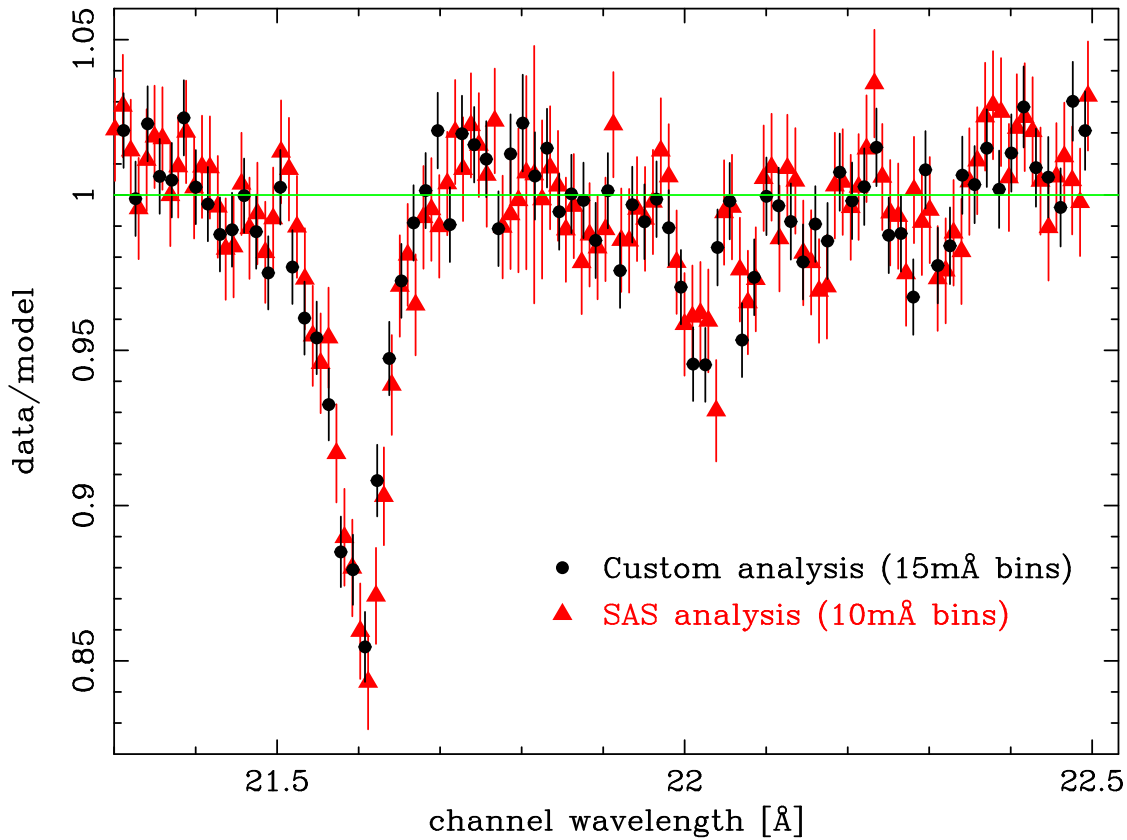}
\caption{Comparison between the spectrum obtained via the XMM SAS analysis
package and our custom software. Binning is slightly different and each 
package identifies slightly different occasional bad pixels. Nevertheless,
the spectra obtained are practically identical.
\label{SAS_custom}}
\end{figure}

The final spectrum obtained in this way, using the SAS, 
appears to be equivalent to 
the spectrum obtained in our custom analysis. This can be seen in
Figure~\ref{SAS_custom}. 

\end{document}